
\tolerance=2000\hbadness=2000
\overfullrule=0pt
\magnification=1200
\baselineskip=12pt
\advance\voffset by 0.4truecm

\parindent=0pt

\def\CC{C\!\!\!\!I}

\font\ninerm=cmr9

\font\twelveit=cmti12

\font\bfeins=cmbx12
\font\bfzwei=cmbx12 scaled \magstep1


\def\abs{\par \vskip 0.3cm \goodbreak\noindent}
\def\Abs{\par \vskip 1.7cm \goodbreak\noindent}
\def\FF{\par\vfill\eject}
\def\IfFF{\FF\ifodd\pageno\else
{\nopagenumbers{\centerline{}\eject}}\fi}
\def\IfPN{\ifnum\pageno=1\else
   {\hss{\ninerm ---  \quad\folio\quad  ---}\hss}\fi}
\def\litem{\par\noindent\hangindent=1.5cm\ltextindent}
\def\ltextindent#1{\hbox to \hangindent{#1\hss}\ignorespaces}
\def\Mittefrei#1#2{\hbox to \hsize{#1\hss#2}}

\def\Ues{\par\nobreak\vskip 0.75 cm\nobreak\noindent}

\def\3{\ss}   

\def\A{{\cal A}}

\def\t#1#2{  {t^{#1}}{}_{#2}  }



\footline={\IfPN}
\abs
\Mittefrei{February 1994}{LBL-35299}
\Mittefrei{}{UCB-PTH-94/03}
\abs\abs
\vskip 1cm
{\baselineskip=16pt{
\centerline{\bfzwei Braided Hopf Algebras and Differential
Calculus$^\ast$}}}

\vskip 1cm
{\baselineskip=14pt{
\centerline{\twelveit }\abs}}
\abs
\centerline{\bf Michael Schlieker$^\dagger$ and $\,$ Bruno Zumino}
\abs
\centerline{\it Department of Physics, University of California}
\centerline{\it and}
\centerline{\it Theoretical Physics Group}

\centerline{\it Lawrence Berkeley Laboratory}
\centerline{\it University of California}
\centerline{\it Berkeley, California 94720}

\vskip 3cm

\centerline{\bf Abstract}
\abs
 We show that the algebra of the bicovariant differential calculus on
a quantum group can be understood as a projection of the cross
product between a braided Hopf algebra and the quantum double of the
quantum group. The resulting super-Hopf algebra can be reproduced by
extending the exterior derivative to tensor products.
\footnote{}{\it $^\ast$ This work was supported in part by the
Director, Office of Energy
Research, Office of High Energy and Nuclear Physics, Division of High
Energy Physics of the U.S. Department of Energy under Contract
DE-AC03-76SF00098 and in part by the National Science Foundation
under
grant PHY-90-21139.}
\footnote{}{\it $^\dagger\,\, $ Supported in part by a

 Feodor-Lynen Fellowship.}

\vskip 1in
\centerline{\bf Disclaimer}
\vskip .2in
This document was prepared as an account of work sponsored by the
United
States Government.  Neither the United States Government nor any
agency
thereof, nor The Regents of the University of California, nor any of
their
employees, makes any warranty, express or implied, or assumes any
legal
liability or responsibility for the accuracy, completeness, or
usefulness
of any information, apparatus, product, or process disclosed, or
represents
that its use would not infringe privately owned rights.  Reference
herein
to any specific commercial products process, or service by its trade
name,
trademark, manufacturer, or otherwise, does not necessarily
constitute or
imply its endorsement, recommendation, or favoring by the United
States
Government or any agency thereof, or The Regents of the University of
California.  The views and opinions of authors expressed herein do
not
necessarily state or reflect those of the United States Government or
any
agency thereof of The Regents of the University of California and
shall
not be used for advertising or product endorsement purposes.
\vskip 2in

\centerline{\it Lawrence Berkeley Laboratory is an equal opportunity
employer.}
\Abs
\FF
{\bfeins 1. Introduction}\Ues
\bigskip
On the superalgebra of the bicovariant differential calculus on a
quantum group\footnote\dag{For simplicity we assume that the quantum
group is of type $A_{n-1}$} $A$ we introduce a Hopf algebra structure
in two different ways. The first approach starts from a set of
commutation relations between the Hopf algebra maps of $A$ and the
bicovariant exterior differential $d$. These commutation relations
yield a consistent super- Hopf algebra structure on the differential
algebra. The resemblance of this structure with the ordinary Hopf
algebra of inhomogeneous quantum groups leads to the second approach.
Starting from the braided Hopf algebra of the Maurer-Cartan

one-forms it is possible to obtain a Hopf algebra by a bosonization
procedure. This Hopf algebra contains the quantum double of $A$ as a
subgroup. By introducing an appropriate projection it is possible to
reproduce the results of the first approach.
\Abs

{\bfeins {2. Hopf structure for bicovariant Differential calculus on
Quantum Groups}}\Ues
\abs
 We first recall the general notions for a bicovariant first order
differential calculus on a quantum group $A$ with multiplication $m$,
comultiplication $\Delta$, antipode $S$ and counit $\epsilon$
introduced by Woronowicz in ref. [Wor].

A bicovariant first order differential calculus is  a differential
calculus

$(\Gamma , d)$ such that $\Gamma$ is a left and right $A$-comodule
and $d$ satisfies :
$$d :\;A\to\Gamma $$
$$d(uw) = (du)w + u(dw) $$
$$\Delta _R(du) = (d\otimes id) \Delta (u) { \rm \;and \;}
\Delta_L(du) = (id\otimes d)\Delta (u)\,, \eqno (1.1) $$
where $\Delta_R$ and $\Delta_L$ are left and right coactions of $A$
on $\Gamma$.
Furthermore we require that $\Gamma$ is a bimodule over $A$ ([Wor]).
The explicit bimodule relations compatible with the coactions on the
space of

one-forms have been studied in terms of $R$-matrices in refs.:[CSWW],
[Jur], [Zum].

In the following we use a left-invariant basis $\Theta$
for the above defined

bicovariant bimodule $\Gamma$. These basis elements $\Theta$ are
expressed in terms of the Maurer-Cartan forms $\omega^i{}_j =
m(S\otimes d) \Delta (t^i{}_j)$ ( $t^i_j$ are the generators of the
quantum group $A$ ) through
$$ {\omega^i{}_j} = {\Theta^k{}_{l}} {\chi^l{}_k} (t^i{}_j) \,, \eqno
(1.2)$$
where $\chi$ are the vector fields on the quantum group $A$ defined

through $du = (\chi^l{}_k \ast u ) \Theta^k{}_{l}$ (compare

 [CSWW], [Jur], [SWZ], [Wor], [Zum]).
The bimodule relations have the form :
$$\Theta^i{}_j u = ((l^{+i}{}_k S(l^{-})^l{}_j)\ast u)
\Theta^k{}_l\,,\eqno(1.3)$$

where $l^{+}, l^{-}$ are the generators of $A^*$ introduced in [FRT].
Using the above relations, the algebra corresponding to the
bicovariant differential calculus can be defined in the following way

$$ \A := \CC<\t ij, \Theta^i{}_j>\bigg/ (I, J, K)\,, \eqno (1.4)$$
where $I$ is the defining ideal of the quantum group $A$, $J$ is the
ideal which

is generated by the bimodule relations (1.3)

and $K$ is given by
$$K := P_r{}^{jl}{}_{ik} \Theta^i{}_j\Theta^k{}_l\,, \eqno(1.5)$$
where $P_r$ are the projectors defined in [CSWW] by the following
equations:
$$ P_r ({\bf 1}- \sigma) = 0\,,r=1,2$$
$$\sigma^{ii'l'l}{}{}{}{}_{jj'k'k} = (l^{+i}{}_k
S(l^{-})^l{}_j)(S(\t{i'}{k'})\t{l'}{j'})\,.\eqno(1.6)$$
On the so defined differential algebra we now obtain a (super-) Hopf
algebra structure starting from the following requirements,
$$ \eqalign{
    d(a \otimes b)& = da \otimes b + (-1)^{|a|} a \otimes db\,,\cr
    [ \Delta , d]& = 0\,,\cr
    [ S , d ]& = 0\,, \cr
    [ \epsilon , d ]& =0\,,\cr
   S(ab) &= (-1)^{|a|\cdot |b|}S(b) S(a)   \,,\cr
    (1\otimes a) (b \otimes 1)& = (-1)^{|a|\cdot |b|} (b\otimes a) ,
{ \rm \;a,\; b\in\A \,are\,homogeneous \,elements\;of\; degree \;|a|,
\;|b| \;.} \cr}\eqno (1.7) $$
The algebra homomorphism property of the comultiplication with
respect to the bimodule relations which generate the ideal $J$
follows directly from the definition of the bicovariant bimodule
$\Gamma$.
The algebra homomorphism property with respect to eq.(1.5) can be
shown as follows: using eq. (1.2) it is easy to calculate the
coproduct, the counit and the antipode for the $\Theta$'s,
$$\eqalign{
\Delta (\Theta^i{}_j) & = \Theta^k{}_l\otimes  S(\t ik)\t lj +
1\otimes \Theta ^i{}_j\,,\cr
\epsilon (\Theta^i{}_j) & = 0\,,\cr
S(\Theta^i{}_j) & = - \Theta^k{}_l(S(S(\t ik)\t lj))\,.\cr}\eqno
(1.8)$$
Inserting this comultiplication in eq.(1.5) and using the definition
of the

projectors from eq.(1.6) together with the bimodule relations one
verifies

immediately that the comultiplication on the forms leaves the ideal
$K$ invariant. All other Hopf algebra axioms can be checked directly.
Writing the comultiplication, the antipode and the counit in terms of
the basis forms $\Theta$ exhibits a close resemblance between the
differential (super-) Hopf algebra and the Hopf algebra of
inhomogeneous quantum groups as introduced in reference [SWW]. It was
shown by Majid [Maj1] that one can reconstruct the Hopf algebra of
inhomogeneous quantum groups through a bosonization procedure from
the braided Hopf algebra of pure translations. In the next chapter we
recall the notions in [Maj1] and then we show that the above defined
differential
Hopf algebra is related to a Hopf algebra which fits in the general
bosonization scheme as defined in refs.:[Maj2][Maj3]. \par

 The procedure defined in eq.(1.7) is applicable to every bicovariant
differential calculus on a Hopf algebra if the differential algebra
has a basis of left invariant one-forms $\Omega^I$ with quadratic
relations $R_{IJ} \Omega^I \Omega^J = 0$ and satisfies
$$R_{IJ} \, ( \Delta _R (\Omega^I) \Delta_L (\Omega^J) + \Delta_L
(\Omega^I)\Delta_R(\Omega^J))=0\,.\eqno(1.9)$$
\Abs

{\bfeins 3. Bosonization of the braided Maurer-Cartan Hopf algebra
}\Ues
Following e.g. ref. [Maj3]  $\cal B$ = $(B,\Delta, S, \epsilon)$
is called a braided Hopf algebra if $B$ is an algebra and the maps

$\Delta : B\to B\otimes B$,  $\epsilon : B\to \CC$,  $S: B\to B$
satisfy the Hopf algebra  axioms with respect to the following
algebra structure in $B\otimes B$:
$$ (a \otimes b )\cdot (c \otimes d) = a \Psi (b \otimes c) d
\,,\eqno(2.1)$$
with $\Psi$ replacing the normal transposition $\tau$ ( the detailed
properties of $\Psi$ are listed in [Maj3]).
The antipode is braided antimultiplicative

$$S(ab) = m \Psi (S(a) \otimes S(b))\,.\eqno(2.2)$$
The algebra $M$ of the Maurer-Cartan forms $\Theta$ with relations
(1.5) can be turned into a braided Hopf algebra using the following
definitions:
$$  \eqalign{
    &\Delta (\Theta )= \Theta \otimes 1 + 1\otimes \Theta =
\Theta_{(1)} \otimes \Theta_{(2)}
\,,\cr
    &\epsilon (\Theta) = 0\,,\cr
    &S (\Theta) = -\Theta.\cr } \eqno (2.3) $$
The Braiding is given by
$$\Psi (\Theta ^i{}_j \otimes \Theta ^k{}_l) = (-1)\sigma ^{-
1}(\Theta ^i{}_j \otimes \Theta ^k{}_l)\,.\eqno (2.4)$$
In the following $\otimes$ stands for a graded tensor product with
standard grading (compare eq.(1.7)).
The braiding $\Psi$ satisfies all the properties listed in [Maj3].
Now we consider the coaction of the Hopf algebra $A$ on the braided
Hopf algebra of the Maurer-Cartan forms. In the following we assume
that the one-forms $\Theta$ transform trivially under the left
coaction and form a representation space for the right adjoint
coaction. Therefore we are able to follow the construction by Majid
([Maj1]). First of all one has to identify the braiding introduced in
(2.4) as being the adjoint representation of some universal structure
defined with respect to the Hopf algebra $A$, i.e. if $X$, $Y$ are
two right comodules of the quasitriangular Hopf algebra  $A$
then the braiding can be written in the form :
$$\Psi (x\otimes y) = y^{(1)}\otimes x^{(1)}{\,}R (x^{(2)} \otimes
y^{(2)})\,,\eqno(2.5)$$

where $\Phi (x) = x^{(1)}\otimes x^{(2)}$, $\Phi (y) = y^{(1)}\otimes
y^{(2)}$ are the right coactions and $R$ is the universal R-matrix of
$A$.
The algebra relations for this generalized semidirect product are
given by:

$$(x \otimes 1)(1\otimes a) = (1\otimes a^{(1)})(x \triangleleft
a^{(2)}\otimes 1)\,,\eqno (2.6)$$
with $x \in X$ and $a \in A$ and
$$x \triangleleft a := x^{(1)} {\,}R (x^{(2)} \otimes a)\,.\eqno
(2.7)$$
In the case of the braided Hopf algebra of Maurer-Cartan forms one
has to be more careful in defining the bosonized Hopf algebra because
applying  eqs.(2.6),(2.7)  in a naive way

would lead to a trivial differential calculus where the scalar
one-form $X$ (see [CSWW]) commutes with all the elements of $A$. Such
a differential calculus does not coincide with the calculus defined
in section 2. Even allowing for some function of $R$ in eq. (2.6)
does not solve the problem because the right transformation of $X$ is
trivial and therefore does not respect the definition of the exterior
derivative on forms. This problem can be solved  by realizing that
the dual Hopf algebra $A^*$ can be embedded as a $\ast$-Hopf
subalgebra in the quantum double $D(A^*)$ of $A^*$ defined by [RS]
(see [CEJSZ], [Jur2], [Pod] for the involution properties):
$$ \eqalign{
    \bf m &= (m\otimes m) \sigma_{12}\,,\cr
    \bf {\Delta} &= R^{-1}_{23} \tau_{23} (\Delta \otimes \Delta
)R_{23} \,,\cr
    \bf {\epsilon} &= \epsilon \otimes \epsilon\,,\cr
    \bf {S}&= R_{21} (S \otimes S)R^{- 1}_{12}\,, \cr
    \bf {R}& = R^{- 1}_{41}R^{- 1}_{42}R_{13}R_{23}\,,\cr
(a \otimes b)^{\dagger} &= R_{21} (b^{\dagger} \otimes
a^{\dagger})R^{- 1}_{21}\,.\cr} \eqno (2.8) $$

( In the following we will use the Hopf algebra ($D(A^*)$, ${\bf
{R}}^{- 1}_{3412}$)). The Hopf subalgebra embedding is given by
$\Delta (A^*)$.
The connection to the bosonization procedure described above is done
using the following identity:
$$ {\bf {R}}^{- 1}_{3412} (S(t)^i{}_j \otimes \t kl \otimes id
\otimes id ) = \Delta (l^{+ i}{}_j {S(l^{-})}^k{}_l)(id \otimes
id)\,.\eqno (2.9)$$
Eq. (2.9) implies that
$$ \Delta_R : B \to B \otimes  D(A^*)^* ,{\rm \; with \;}\,\Delta_R
(\Theta ^i{}_j )= \Theta^k{}_l \otimes S(t)^i{}_k \hat {\otimes} \t
lj =
\Theta^{(1)} \otimes \Theta^{(2)}\,, \quad\eqno(2.10)$$

where $a \hat {\otimes}b$ is the bicrossproduct in $D(A^*)^*$ and
satisfies
$$(a \hat {\otimes} b)(c \hat {\otimes} d) = {\bf m}(R_{32} \ast (a
\otimes b \otimes c\otimes d) \ast R^{-1}_{32})\,.\eqno (2.11)$$
 On the other hand eq.(2.9) also shows that adopting this convention
is enough to reproduce the bicovariant differential calculus
described in section 2 because the half representation of the
universal R-matrix defined in eq.(2.9) lies already in the Hopf
algebra embedding of $A^*$ in $D(A^*)$.
 Therefore it is not necessary to extend the coproduct on $A$ to a
mapping to $A \otimes D(A^*)^*$.

\abs
The coproduct of the Maurer-Cartan forms $\Theta$
in the bosonized Hopf algebra $M {\scriptstyle> \!<\!\mid } D(A^*)^*$
is defined through ($\bf 1$ is the unit in
$D(A^*)^*$)\footnote\dag{For simplicity we use the same symbol for
the coproduct in $D(A^*)^*$ and $M {\scriptstyle> \!<\!\mid }
D(A^*)^*$.}
$$ {\bf {\Delta }}(\Theta \otimes {\bf 1}) = (\Theta ^{(1)}{}_{(1)}
\otimes {\bf 1}) \otimes ( 1 \otimes \Theta ^{(2)}{}_{(1)})(\Theta
_{(2)} \otimes {\bf 1})\,.\eqno (2.12)$$
Using (2.5), (2.6) and (2.9) we obtain the following expression for
the algebra relations in $M {\scriptstyle> \!<\!\mid } D(A^*)^*$

$$ \eqalign {
(\Theta ^i{}_j \otimes {\bf 1}) (1 \otimes a {\hat {\otimes}} b)&= (1
\otimes  a_{(1)} {\hat {\otimes}} b_{(1)} )( \Theta ^k{}_l \otimes
{\bf 1})\,\,{\bf {R}}^{- 1}_{3412} (S(t)^i{}_k \otimes \t lj \otimes
a_{(2)} \otimes b_{(2)})\,\cr
          &= (1 \otimes  a_{(1)} {\hat {\otimes}} b_{(1)} ) ( \Theta
^k{}_l \otimes
{\bf1})\,\,(l^{+i}{}_k{S(l^{-})}^l{}_j)(a_{(2)}b_{(2)})\,.\cr}\eqno(2.
13)$$
As already pointed out earlier, eq.(2.13) shows that the commutation
relation

do not distinguish between elements from the quantum double of $A$
and $A$ itself. Therefore it is possible to introduce the following
$\ast$-homomorphism
 from the quantum double to the original
Hopf algebra $A$
$$ \varphi :  D(A^*)^* \rightarrow A , {\rm \; with \;}\,\varphi ( a
{\hat {\otimes}} b) = m (a {\hat {\otimes}} b)\,.\eqno(2.14)$$
The homomorphism property can be verified directly using the
definition of the

product in the quantum double (2.11).

The application of the projection $\varphi$ on $M {\scriptstyle>
\!<\!\mid } D(A^*)^*$ (where $\varphi$ is the identity on $M$ )

 yields the Hopf algebra defined in section 2 through eq.(1.7).
Therefore we obtain the following equivalence:

$$(M {\scriptstyle> \!<\!\mid } D(A^*)^*, {\bf{\Delta, S, \epsilon}})
\buildrel {\varphi}\over \longrightarrow (\A,\Delta, S,
\epsilon)\,.\eqno(2.15)$$

\FF
{\bfeins References}\Ues
\litem{[CEJSZ]} Chryssomalakos, C.,  Engeldinger, R.,  Jur\v co, B.,
Schlieker,

   M.,  Zumino, B.: Complex Quantum Enveloping Algebras as Twisted
Tensor

   Products. Preprint LMU-TPW 93-2
\litem{[CSWW]}  Carow-Watamura, U., Schlieker, M., Watamura, S.,
Weich, W.:
   Bicovariant differential calculus on quantum groups $SU_q(N)$ and
$SO_q(N)$.

   Comm. Math. Phys. {\bf 142}, 605 (1991).
\litem{[FRT]} Faddeev, L. D., Reshetikhin, N. Yu., Takhtajan, L. A.:

   Quantization of Lie groups and Lie algebras.
   Algebra and Analysis {\bf 1}, 178 (1987).
\litem{[Jur1]} Jur\v co, B.: Differential calculus on quantized
simple
   Lie groups. Lett. Math. Phys. {\bf 22}, 177 (1991).
\litem{[Jur2]} Jur\v co, B.: More on quantum groups from the
quantization point

    of view. ASI TU preprint Jan. 1993.

\litem{[Maj1]} Majid, S.: Braided momentum in the q-Poincare group.
   J. Math. phys. {\bf 34}, 5 (1993).
\litem{[Maj2]} Majid, S.: The Quantum Double as Quantum Mechanics.
Preprint

  DAMTP/92-48.
\litem{[Maj3]} Majid, S.: Cross products by Braided Groups and
Bosonization.
  Preprint DAMTP/91-11, to appear in J. Algebra.
\litem{[Pod]} Podle\'s P.: Complex quantum groups and their real

   representations. RIMS 754 (1991).

\litem{[RS]} Reshetikhin, N. Yu., Semenov-Tian-Shansky, M.A.: Quantum
R-matrices and factorization problems. J.G.P. {\bf 5} 533 (1988).
\litem{[SWW]} Schlieker, M.,  Weich, W., Weixler, R.: Inhomogeneous
Quantum

Groups. Z.Phys. C-Particles and Fields {\bf 53} 79 (1992).
\litem {[SWZ]} Schupp, P., Watts, P., Zumino, B.: Bicovariant Quantum
Algebras and Quantum Lie Algebras.  Comm. Math. Phys. {\bf 157}, 305
(1993).
\litem{[Wor]} Woronowicz, S.L.:
   Differential calculus on compact matrix pseudogroups (quantum
groups).
   Comm. Math. Phys. {\bf 122}, 125 (1989).
\litem{[Zum]} Zumino, B.: Introduction to the Differential Geometry
of quantum groups, in Mathematical Physics X, K. Schm\"udgen Ed.
Springer 1992.

\vfill
\eject
\end